\documentstyle[12pt]{article}
\title{PROBING THE PHYSICS OF AGN -- A SUMMARY}
\author{R. D. Blandford\\Caltech}

\def\go{
\mathrel{\raise.3ex\hbox{$>$}\mkern-14mu\lower0.6ex\hbox{$\sim$}}
}
\def\lo{
\mathrel{\raise.3ex\hbox{$<$}\mkern-14mu\lower0.6ex\hbox{$\sim$}}
}

\begin{document}
\maketitle
\begin{abstract}
A summary critique is presented of the topics discussed at a workshop entitled 
``Probing the Physics of Active Galactic Nuclei by 
Multi-wavelength Monitoring'' held at Goddard Space Flight Center
21-23 June 2000. Particular emphasis is placed on the larger astronomical 
context of research on active galactic nuclei. An important goal is 
to use multi-wavelength monitoring 
to produce ``engineering drawings'' of how gas flows
in the vicinity of a black hole, as well is in the associated accretion 
disk inflows and jet outflows.  Some of the clear choices 
that have to be made before formulating a model of a particular
class of source are outlined. Future observational possibilities are
briefly summarized.  
\end{abstract}
\section{Introduction}
There is an analogy, which is sometimes drawn, between the
unfolding of our understanding of the structure of atoms and of galaxies.
In 1909, Geiger \& Marsden, working under Rutherford's direction, discovered that
atoms contain compact, positively charged nuclei occupying $\sim10^{-15}$ of their
volume. The path through Bohr theory, non-relativistic quantum mechanics, spin,
the exclusion principle, relativistic quantum mechanics and quantum
electrodynamics is well-trodden by physics and astronomy students. 
In less than two decades physicists had 
assembled an elegant and precise set of rules for describing atoms that has
only changed by quantitatively tiny (though still immensely important) 
corrections. One year prior to the work of Geiger \& Marsden,
Edward Fath, then working at the
Lick observatory, obtained the first evidence that galaxies too contain
compact nuclei, in this case, as we now know, occupying $\sim10^{-30}$ of their volume.
However, it took astronomers over fifty years to associate these nuclei with 
black holes and another twenty years to assemble the evidence that 
convinced doubters that black holes were generically present in most normal
galaxies.

My reason for bringing this up is to take issue with the title. The understanding of 
atoms required new and fundamental physical principles. It is probably the case that 
no such principles are needed to describe AGN. The main ingredients, Newtonian
physics, electromagnetic theory, atomic astrophysics, hydrodynamics, radiative transfer, QED etc are
so well tested that it is hard to imagine that AGN observations could
cause any of them to be cast into doubt. Even general relativity, although only tested in 
the weak field regime, is so strongly ensconced in our discipline that the Kerr metric 
will continue to be used to set the stage for black hole astrophysics unless some 
truly compelling, quantitative discrepancy is reported. The one subject where we really 
are quite ignorant of many relevant,
fundamental principles is magnetohydrodynamics, but this is probably
not what the organisers had in mind! (I would comment that, with the advent of high
dynamical range 3D MHD simulations and with the availability of superb observations of the 
solar corona and terrestrial magnetosphere, the prospect for understand how 
magnetic reconnection, strong hydromagnetic shocks and so on actually operate are 
pretty good and AGN studies ought to be the grateful beneficiary.)

In fact, I really think what we need to be discussing is the \emph{engineering} of AGN. Quasars,
Seyfert galaxies and so on are sleek machines. They take gaseous fuel, organize 
it into a disk and, when conditions are right, convert its mass into
radiant energy and its exhaust into jets with an efficiency that would win 
accolades in the automotive
industry. However, despite the growth in our observational understanding over the 
past decade, that has made AGN studies much more quantitative, we still have not had 
a good look ``under the hood''. We really do not have a common, agreed description of how 
AGN function -- what roles are played by the hole, the disk, the stars and so on and how the
different elements, emission line regions, narrow line regions, jets, winds, disk coronae
are geometrically arranged.
This is where the techniques that have been discussed over the past three days
come in. What they promise is to answer some of these outstanding questions by direct 
measurement. 

What I have been asked to do is to provide a critique of the field as it
has been described as opposed to a traditional conference summary. I have tried to be 
provocative and raise, in advance,
some of the questions that will be asked of the large observing and mission 
proposals on which the future of this field depends. I hope I can be excused preparing
a bibliography, referring instead by name to the accompanying articles where contemporary
references can be found.  
\section{The Bigger Picture}
Most of this meeting has been concerned with attempts to understand AGN variability 
in the optical and X-ray emission lines and in relativistic jets (Peterson). 
In practice, these
are three rather distinct research communities, at least observationally and I am not sure
how much they have to learn from each other in terms of technique. Where there is much
common ground, and where the discussion has been relatively parochial, is in understanding
how AGN engineering fits into a much larger astronomical enterprise. I believe that the context
under which AGN are studied is changing quite rapidly in several different ways.
\subsection{Galaxy Formation}
This is an exciting time in the study of galaxy formation. Although theory plays a crucial role
here, this is fundamentally an empirical business. HST and 10m class telescopes, 
used in conjunction, are enabling detailed observations of galaxies that were formed when the 
universe was less than three billion years old and finding galaxies out to $z>5$. Quasar
observers, not to be outdone, are matching them in redshift. This is significant.  As far as
we can tell, the formation of a central nuclear black hole is an essential concomitant of galaxy  
formation. There is a growing acceptance of the notion that active nuclei could
have provided a crucial feedback that 
limited the growth of galaxies, long before most of their stars were born. We do not know 
if galaxies form from the inside out or the outside in. Furthermore, as I have 
speculated elsewhere, it is possible that massive black holes make elliptical galaxies 
(through preventing disk formation) rather than 
the other way around as is generally supposed. Clearly, such speculations cannot be addressed
until we have a better understanding of how black holes grow and this will require
being able to mass them at large redshift from their spectra and luminosities. This, in
term, requires that we understand how they work.

To be more specific, most effort has been concentrated on studying local AGN, where
our diagnostics are the best and where some promising correlations
are emerging between the mass of the hole and the properties 
of the surrounding galaxy [Wandel]. 
What we really need to do now is to focus on some 
attainable goals like measuring the mass accretion
rate (in units of the Eddington rate), the spin angular frequency of the hole (in units of the
inverse mass) and relating these to the 
character of the immediate environment (both stellar and gaseous). We then need to understand
the associated, radiative efficiencies and the strength and type of outflow. 
If techniques like reverberation mapping can supply the rules for measuring these quantities
[Wandel, Netzer], then 
we would be in a very good position to apply them to distant galaxies,
where we cannot make dynamical measurements of the hole mass. This allows us to 
undertake quantitative, demographic studies of young galaxies
and their nuclei
to replace the guesswork with which we have to make do at the moment. 
\subsection{Genesis of Black Holes}
Just as it has proven far more difficult to explain how stars form as opposed 
to how they function, so we must expect discussions of the origin 
of black holes to be relatively speculative.  Are black holes
formed on a dynamical timescale with masses larger than $\sim10^6$~M$_\odot$
or do they grow critically from much smaller masses {\it in situ}?  Alternatively, are they formed
as an end product of intergalactic ``Population III'' stars at even earlier
epochs? Again, understanding the rules which determine how fast black holes grow as a function of 
their mass and mass supply and environment
will be necessary to rule out false accounts. 
\subsection{Ionization of the Intergalactic Medium}
A crucial question is ``When and how was the intergalactic medium ionized?''. The answer 
almost surely
involves both hot young stars and quasars, but it is proving hard to be quantitative.
The particular importance of quasars is that they produce
hard, penetrating photons. In order to test this quantitatively, we need to
know how to interpret the observations of high redshift quasars, take out the beaming effects and 
estimate their average ionizing flux, when we cannot determine this directly
on account of the very absorption that we are trying to explain.
\subsection{Scale-Free Accretion}
This raises the question of the scaling of black hole properties with mass; in particular
the comparison of stellar mass black hole accretors with massive black holes in AGN. On
the one hand, the dynamics and two body processes associated with optically thin accretion
are essentially scale free (Quataert, Narayan). However, what is not scale-free 
is the effective temperature of the radiation when it is  
absorbed and re-radiated. We do not really have a good understanding 
of how much this influences the dynamics of accretion.
Studying matched Galactic and AGN systems  (Livio) should 
be very helpful. In addition, if there is a
large population of intermediate mass black holes as suggested by recent 
Chandra observations then we shall have another important point of comparison 
\subsection{The Advantage of Monitoring X-ray Binaries}
This leads us naturally to the question raised by a couple of speakers.
``Why put so much effort into monitoring AGN when you can get so much better
sampling of Galactic black hole transients?'' In the most naive interpretation,
all timescales should scale with mass and a year in the life of a quasar is lived in 
a second by an X-ray binary. Isn't this the best way to find
out how AGN work? The RXTE observations of GRS 1915+105 provide graphic 
illustration of the quality of the data that can be procured in a short time.

The best answer that I can give to this question, and the best 
defense of a major 
effort in AGN monitoring is that, despite the appeal of the scaling laws,
X-ray binaries and AGN exhibit important differences. The former do not exhibit
ultrarelativistic jets or prominent broad emission (or absorption) lines.
In addition, X-ray binaries do not, as yet, display broad iron lines in the 
X-ray part of the spectrum. Making the inverse comparison, we haven't really 
found a massive counterpart to SS433. 
The environments are also quite different - a single, binary companion
versus a star cluster. 

Comparison of the two types of black hole will probably turn out to be 
pretty important as we sort out the details.  
\subsection{$\gamma$-ray Bursts}
Another under-represented point of comparison is $\gamma$-ray bursts. After thirty years hard work,
radio astronomers
have assiduously built up a picture of relativistic jets and tentatively
pushed up the Lorentz factors to 15. More recently, the gamma ray burst community has 
recapitulated much of the history of the study of compact radio sources and added
several new ingredients, the most important of which is the deduction that 
the Lorentz factors initially exceed several hundred. They too have found some 
evidence that bursts may be beamed. To me, the similarities between these 
two classes of object are more impressive than the differences. The mysteries
of putative $\gamma$-ray jets provide a larger justification for the study of their
counterparts in AGN.  
\subsection{The Intergalactic Infrared Background}
My last example of the wider importance of AGN monitoring concerns TeV observations
of blazars. As is well known TeV $\gamma$-rays are susceptible to absorption by infrared 
photons with the production of electron-positron pairs. As such, they provide a splendid probe
of intergalactic infrared emission and, by extension the rate of star formation
in the Universe. Observations with decreasing energy can probe to greater redshifts. However,
to make these measurements, usefully quantitative, we have to understand what is happening
in the sources.
\section{How does the Gas Flow, Blow and Glow?}
As a pragmatic matter, it is sensible to accept ``old'' physics uncritically and just focus
on trying to understand the dynamics of the various AGN components -
disks, jets, clouds and so on. Impressive progress was reported here.
\subsection{Round the Clock Surveillance}
It requires very little thought to write a proposal to monitor the flux from a set
of variable sources with the hope that something useful will be learned as a consequnce.
Even if this is ever a sound strategy, it is certainly not acceptable for AGN where 
we know far too much to be excused from devising an optimal strategy for using the 
considerable amounts of telescope time that are needed. sampling is crucial. Early 
variability studies, based on overestimates of the size of the emission line region,
were often seriously undersampled. (Some contemporary studies,
despite heroic efforts, remain so [Balonek].)  In addition, it is necessary to observe for
long enough that the characteristic variation and light crossing timescales are
well covered. It was therefore a welcome surprise that a study of quasars [Kaspi]
should have yielded such promising results.
The far more analytic approach, advocated here 
by Horne, Peterson and others, will surely be necessary to justify a large space mission.

In a quite different approach, recent ideas in describing self-organized criticality
were applied to AGN light curves (Mineshige, Kawaguchi). From a physics perspective, these ideas are very 
attractive, but as an astronomer, I personally hope that their application here turns out to 
be unsuccessful
because if they are right, they give us little hope for using variability 
for understanding how black holes accrete.
\subsection{Microvariability}
We heard impressive reports of high precision photometry at the 0.05$^m$ level [Miller,
Ferrara]. This is apparently detectable in 85 percent of the radio-loud quasars
but a smaller fraction of their radio-quiet counterparts. There seem to be associated 
color changes. Although several models involving disks and
outflows, were discussed, it will be quite hard to decide
between them unless the variation can be tied to similar changes in other parts of the
electromagnetic spectrum.
\subsection{In, Out, Round and About}
The first task in reverberation mapping is surely kinematic - to determine 
the state of motion of the emission line gas. 
Four general types of flow have generally been distinguished and advocated. 
\begin{itemize}
\item We may be witnessing gas as it flows onto the hole [Welsh]. This is surely
the most natural expectation.  
\item Most of the illuminated
gas may be in a wind, which is surely present in the case of the broad
absorption line quasars [Everett]. 
\item The emission line clouds may be part of the disk. However,
this can be quite complicated. The disk may be shadowed by an inner torus
or warped so that illumination can be quite irregular [Shields].
\item The motion may be so chaotic that it is most useful to describe as 
locally isotropic at all points around the continuum source [Fromerth, Collier]. 
\end {itemize}

Distinguishing these four generic 
possibilites unambiguously has not, been easy, although considerable progress has been reported
here. I think it safe to conclude that the accretion flow is not simple!
\subsection{Some Systematics}
Enough AGN have been monitored that patterns are starting to emerge.
Some of these were expected;
others are more of a surprize [Alexander, Sun, Mathur, Korista, de Salamanca]. 
\begin{itemize}
\item Certain optical emission lines are proving to 
exhibit far stronger echo responses than others. There does not appear 
to be a very good understanding of why this is so.
\item The ionization parameters that are measured, particularly in objects like NGC 5548,
turned out to be 
nearly an order of magnitude larger than anticipated [Netzer]. 
This makes life more interesting for 
the observer.
\item There is clear evidence for ionization stratification with the higher ionization
states being located closer to the continuum source. This is even reported in the UV
with the OVI being co-spatial with the HeII [Kollatschny]. 
\item The ionization parameter seems to decrease with increasing 
luminosity. 
\end{itemize} 
\subsection{Narrow Line Seyfert 1 Galaxies}
There has been much attention paid lately and at this meeting to the so-called
Narrow Line Seyfert 1 Galaxies [Leighly, Turner, Edelson]. In 
addition to narrow emission lines these also 
exhibit warm absorbers with high ionization states and steep X-ray spectra.
It is tempting to suggest that these may be the low luminosity counterparts of
broad absorption line quasars. 
\subsection{Dynamics of the Emission Line Gas}
Frequently this gas is characterized as being in the form
of tiny clouds moving hypersonically ($M>1000$) through a dense, confining medium.
This is absurd. Nonetheless this model provides a fair representation of the observed emission 
line strengths. From a dynamical perspective, it makes much more sense to regard the 
gas as part of a flow, possibly driven by radiative or magnetic stresses (Bottorff). 
The challenge is to reconcile the radiative description with the dynamical.
\subsection{The Inner Disk}
One of the most impressive astronomical results over the past few years was the measurement 
of broadened iron lines from the several low power Seyfert galaxies and LINERs
[Nandra, Reynolds]. 
These have been interpreted as arising through fluorescence
from thin accretion disk 
photospheres illuminated by an X-ray continuum source at high latitude.
Alternative explanations have been patiently explored, but seem generally 
unconvincing, although the influence of Compton scattering may yet turn out to 
be unimportant.
This is all fitting in well with an encouragingly ordered view of X-ray emission from 
this type of AGN. Specifically, this model requires there to be an extended,
active corona maintained at a temperature $\sim100$~keV, presumably by magnetic activity 
derived from the underlying accretion disk.  The
hot electrons scatter soft photons from the disk producing a power law X-ray spectrum, half
of which irradiates the disk producing the characteristic reflection spectrum as well as 
the iron lines. This X-ray emission exhibits fast and slow variation. It is tempting 
to attribute the former to magnetic activity and the latter to variations in the soft photon
flux. If this model is correct, then the observed X-ray spectrum ought to turn 
over above $\sim100$~keV as now appears to be the case.

There were some serious concerns raised here about the reliability of the 
observed line profiles 
and their apparent lack of variation (Nandra, Reynolds, 
Edelson). However if we look on the positive 
side then these observations have opened the door to a much richer spectroscopy of accretion 
disk surfaces and to reverberation mapping which offers the promise of filling in 
many of the details of gas flow around black holes.   
\subsection{The Outer Disk}
There were also encouraging reports of patterns in the behavior of gas in the outer disk
with variability timescales varying with wavelength $\propto\lambda^{4/3}$ (Peterson) as well
as slower variation attributable to disk dynamics [Maoz, McHardy]. Fairall 9 
and Mrk 744 exhibit infrared 
reverberation presumably associated with hot dust located decently outside the sublimation
radius [Nelson,Oknyanski]. Probing the radial variation of the dust temperature 
is becoming increasingly promising.
\subsection{What Happens to Binding Energy?}
Narayan and Quataert presented two very clear summaries of the prodigious number of papers 
written on ADAF models of slow accretion onto black holes. 
I cannot pass up the opportunity to present, once gain,
a contrary view. The simple ADAF models are 
predicated on three dynamical assumptions, that gas 
cannot radiate, that the flow is mass-conservative and that the flow
is stationary. On this basis it is concluded that the gas supplied crosses 
the event horizon advecting the released binding energy with it.
The first hypothesis
depends upon microphysical considerations that have always been a bit problematic but 
it has certainly become more plausible recently that it is correct and is definitely
worth entertaining. The second hypothesis is the one with which I have the greatest
difficulty. Put simply, when gas accretes rotationally onto a black hole, there must either 
be a torque that transports angular momentum {\it out} through the disk or one that carries 
it {\it up} to the disk surface and off in a hydromagnetic wind. In the former case
there is an unavoidable  transport of energy
out through the disk from small radius to large radius. (It is this effect that is responsible
for accreting gas in a radiative disk 
emitting three times the binding energy that it releases.) 
The transported energy must go somewhere and if the disk cannot radiate, the energy
must be carried off in a wind. I do not believe that it is possible for it 
all to cross the event horizon.
Under most circumstances, we expect that there is a powerful
wind although this is not necessary. The energy may be extracted by magnetic torques.
Either way, the flow does
not look like the original ADAF proposal. 

To go beyond these general principles requires understanding the microphysical details. 
One longstanding possibility, borne out by recent numerical simulation, is that the flow becomes 
convective and establishes a net mass inflow that increases with radius. However, this does
not solve the problem. Most of the mass supplied must still be lost.
The differences are not small. In the case of the Galactic center, 
we might estimate that the rate of gas {\it supply}
is $\sim10^{22}$~g s$^{-1}$. In the ADAF model, almost all of this
goes down the hole with a radiative efficiency $\sim10^{-7}$. By contrast, if there is a
wind, simple mass loss models suggest that the rate of gas {\it accretion} is 
only $\sim10^{18}$~g s$^{-1}$ - one proton out of ten thousand altruistically sacrifices 
itself to enable the remainder to escape - and the radiative efficiency is $\sim10^{-3}$.

An alternative possibility is that the third assumption is incorrect and the flow is cyclical, 
building up a reservoir until the gas can accrete radiatively. If this is happening, then we 
should see evidence for past radiative phases, a possibility that can be dismissed in the case
of the Galactic center.

There is an additional concern. The basic ADAF models suppose that the electrons maintain
a Maxwellian distribution function with the minimum temperature permitted by by Coulomb 
interactions. However, we are probably dealing here with a trans-relativistic, trans-sonic,
trans-Alfv\'enic rapidly shearing, hydromagnetic flow. Whenever we observe mildly 
shearing hydromagnetic flows in the heliosphere, they are accompanied 
by particle acceleration. In the case of accretion onto a black hole,
a comparatively small admixture of MeV electrons 
can boost the radiative efficiency by orders of magnitude. Furthermore, in the case of the Galactic
center, the X-ray spectrum changes from flat (free-free) to steep (nonthermal). 
Another way of expressing this puzzle is to observe that 
then the assumption that the energy dissipated is channelled into the ions, is in stark contrast
with the assumption made in discussing X-ray emitting coronae where we require that
electrons are efficiently heated.
It is possible that the character of the dissipation depends upon 
the $\beta$ of the plasma, being low in a corona and high in a disk.
This  deserves further study.
\section{Jet Lag}
\subsection{The Diversity of Jets}
This meeting has also been concerned with the behavior of jets. We still lack a 
comprehensive and generally accepted theory of their formation, collimation and radiative 
properties. What is increasingly apparent is that bipolar outflow is a very common
accompaniment of accretion (Livio). Indeed, it may even be necessary to carry off
surplus angular momentum. Having said this, I am somewhat mistrustful of assertions that 
one mechanism will fit all jets. This is not the case for stars, for 
supernovae, or for X-ray or $\gamma$-ray bursts, although all mechanisms have to 
satisfy fundamental conservation laws for mass, momentum, angular momentum and, as I 
have just emphasized, energy. Blazar etc jets are ultrarelativistic, Seyfert jets are not
yet both originate from accretion disks around black holes. Some outflows escape
from radiation-dominated environments where the magnetic stresses might be comparatively 
weak; others probably emerge from hot, optically thin plasmas which may be strongly
magnetized. The former may well be poorly collimated and subrelativistic, like the Seyfert 
outflows and perhaps some of the Galactic sources. The latter may be the ultrarelativistic
``superluminal sources'' and subject to quite different dynamical rules.
\subsection{X-ray Jets}
Chandra, with its arcsecond angular resolution is proving to be a boon to this subject.
The first X-ray jets to be reported, Cen A, 3C 273 and 3C371 etc, are 
suprisingly intense and well-collimated. They show 
X-ray all along their lengths, strongly suggesting that particle acceleration can proceed 
efficiently without relying on occasional strong shocks.

Even more surprising are the jets that we did not expect to find, specifically those
associated with the Crab and Vela pulsars. These are presumably formed without the 
assistance of accretion disks. Perhaps the common element is not a disk, but a spinning source
of organized, poloidal field.
\subsection{Blazars}
Although high frequency VLBI can probe the jets in a few selected sources, like M87
down to size scales $\sim100m$, most blazars cannot be so well-probed, especially
after de-projection.  We must therefore rely on indirect methods including,
especially, multi-wavelength monitoring to understand their
structure. It is to be hoped that these will 
ultimately lead to a correct identification of the source of the high frequency emission.

We have learned a lot observationally [Sambruna, Marscher, Steinle, Marchenko-Jorstad]. 
We are mostly confident 
that these source are 
embodiments of the traditional ``inverse Compton catastrophe'' within a relativistic
outflow. A broad low frequency hump is apparent associated with synchrotron radiation,
At high frequencies, there is a larger, more luminous hump, peaking in the $\gamma$-rays.
There are several proposals as to how these humps are produced. The first question is
the origin of the soft photons that are scattered. 

The traditional answer is that they are
the synchrotron photons from the low frequency hump, presumably emitted at the same 
radius in the jet outflow as the $\gamma$-rays with both components being Doppler-beamed
by the relativistic outflow.  However, an alternative possibility is that the soft photons
have an external origin. Specifically, we expect that the black hole will have a luminous 
accretion disk and that a fraction, perhaps ten percent, of these photons will be scattered
by the outer disk, emission line clouds or the general intercloud medium at the emission 
radius. As the inverse Compton power of an ultrarelativistic 
electron of energy $\gamma m_e c^2$, measured
in the blazar frame, is $\propto\gamma^2<(1-\cos\theta)^2>$, where $\theta$ is the angle between
the direction of the electron motion and the photons. There is clearly a radiative
advantage in external scattering models. Correlating the variability in the synchrotron bands 
is probably the best way to decide between these two general models. We will 
probably have to wait for GLAST to be sure, (although I think that current indications
favor the self-Compton model, in most cases).

This is not the only issue. We also need to understand where and how the relativistic
electrons are accelerated. There is good circumstantial evidence, and even
stronger theoretical expectation that the jet power derives from close to the black hole.
If so, we must reject what is perhaps the most natural possibility -- that it be released
as a pair plasma. This is because the radiative (and momentum) losses close to the hole are
inevitably catastrophic and the outflows could never maintain the ultrarelativistic 
speeds that are directly observed by radio astronomers at much greater radii, where there
is no objection to the plasma being pair-dominated and where there is some indication
that it actually does have this character. There must 
be an alternative carrier of momentum close to the black hole.  The two candidates
are relativistic protons and Poynting flux. (In the case of relativistic  
protons, though, it is hard to see how tightly collimated 
beams could be formed without strong magnetic field
being present and magnetic energy transport being at least competitive with particle transport.)
If the energy flux is electromagnetic 
from the start, organized, rotational kinetic energy is 
tapped by electromagnetic field and transported 
to large distance where it is used up in electron (and positron) acceleration.This is what 
happens in the case of pulsars and, arguably, $\gamma$-ray bursts. 

It is not clear if all this conversion of Poynting flux to particle energy happens 
at a single location -- the one zone model -- or if the dissipation occurs continuously
along the jet.
What is clear is that the highest energy $\gamma$-rays must be created outside an
energy-dependent ``gammasphere'' where the optical depth to pair production
on the ambient soft UV/X-ray background falls to unity. This suggested the possibility
that the emission always originates from the gammasphere. On this basis, it was predicted 
that the higher energy $\gamma$-ray emission would vary more slowly than lower energy 
emission and might exhibit lags, in contrast to what is expected from the
one zone model. (One problem with this explanation is that the cooling pairs, both primary and 
secondary, will emit more X-rays than are observed.)
There is as yet not much evidence for these effects, but again, the data 
are sparse. Sorting all of this out is an exceelent goal for future multi-wavelength 
monitoring.

Another key question is how to model the outbursts. Are they best described 
as essentially stationary jets perturbed by traveling shock fronts that 
increase the emissivity, or should we think of them more as independent explosions
occuring in a previously evacuated channel? Tracing circuits in 
the $S-\alpha$ plane looks like a particularly useful way to address the 
geometrical relationships indirectly [Sambruna, Marscher].
The difference is not just important 
dynamically but also affects the radiative properties.            
\subsection{Jet Velocity}
What is the jet velocity? The reason why this is so crucial is that
relativistic beaming is such a powerful amplifier. The volume emissivity
of a stationary flow is $\propto{\cal D}^{2+\alpha}$, where ${\cal D}$ is the Doppler factor 
and $\alpha$ is the spectral index. (In computing the flux from a moving 
component, an extra power of ${\\cal D}$ must be included.)
This means that relatively
insignificant parts of the source that are directed towards us may dominate the 
observed intensity. 

Now it is quite unlikely that the jet is characterized by a single speed.
There are good reasons for it to accelerate and decelerate along its length and 
even more reasons for it to establish a transverse velocity profile.
Furthermore, the parts that we see are probably shock waves and this introduces
additional kinematic complexity.
In particular, the direction and the speed of the emitting plasma behind a shock front
must differ from the kinematic speed of the shock. Higher angular resolution 
VLBI observations will be necessary to sort all this out in individual sources.

Most models have assumed, adopted or argued for jet Lorentz factors
$\gamma\sim10$. However the observation of intraday variability, and especially
the direct observations of the source sizes are suggesting that the Lorentz factors 
could be much larger. (If we can measure the source size, then we have a direct determination
of the brightness temperature which is boosted by a factor $2\gamma$
from its value in the comoving frame which, in turn,
cannot be much larger than the traditional 
inverse Compton value of $\sim10^{12}$~K; hence the minimum Lorentz factor.) Now, 
it might be thought that the solution is just to keep increasing the Lorentz factor. After
all values of $\gamma\sim300$ are routinely invoked in $\gamma$-ray burst models. However,
this causes a problem with blazars because when the Lorentz factor becomes too large,
typically larger than $\sim30$, the radiative efficiency becomes unreasonably small.
The observations have not yet driven us to this conclusion, but they may soon do so. If so,
the best outcome may be that we should abandon synchrotron radiation and turn to 
some coherent process like a cyclotron maser, that might be pumped behind a
shock front. We are, however, not allowed to adopt unlimited brightness temperature in the 
emission region because the radio waves have to escape and a relatively small quantity 
of plasma renders the emission susceptible to nonlinear effects like induced Compton scattering
and stimulated Raman scattering. I suspect that VLBI circular polarization observations
may be crucial in determining what is happening.
\subsection{TeV Emission}
One of the most pleasing developments in high energy astrophysics in recent years is that 
rapidly (as short as half an hour)
variable TeV emission is detectable from closeby blazars.
This places a lower bound on the emission radius inorder that 
the optical depth to pair production be less than unity.
If there is a component of infrared/optical emission
at wavelength $\sim2\theta^2\mu$ with luminosity $L=10^{43}L_{43}{\rm erg s}^{-1}$,
making an angle $\theta$ with the jet, then
$\theta<3^\circ(L_{43}/t_{{\rm hr}})^{-1/4}$. This is an important constraint which limits the 
amount of external irradiation and places a lower bound on the jet Lorentz
factor. Future TeV variability studies, coupled with studies of the unbeamed parent population 
-- source like  M87 -- should be very interesting. 
\subsection{Spin and Jet Formation}
As I emphasized, there is as yet no widely accepted theory of jet formation. However, there 
is much more interest now in this question, both observationally and theoretically.  
It seems to be generally acknowledged on observational grounds that AGN jets
arise fairly close to the central black hole and that there are some 
sources (eg M87) where they cannot be confined and collimated by 
gas pressure alone and where the jet luminosity almost certainly exceeds the bolometric
luminosity of the nucleus and the Eddington luminosity by several orders of magnitude
so that radiation pressure is also irrelevant.
(This may not be true of all jets, but it at least defines a class of sources that 
we can hope to understand.) This leaves magnetic field as the most likely agency for launching
and sustaining jets, a notion which is very welcome theoretically because of the discovery
of powerful, dynamical MHD instabilities that rapidly build strong field in accretion disks.
However, there are interesting alternatives including proton cascades
[Falcke] that have been proposed.

There are several choices that have to be made before making a model. 
Firstly there is the choice of the prime mover. Does jet power originate from the released
binding energy of gas as it accretes through a 
disk [Falcke] or does it come from the spin of a black hole?
The former is sometimes argued to produce more power than the latter, which is probably 
the case for a thin disk, 
though not necessarily for a thick one. The latter, however, is more likely 
to lead to an ultrarelativistic outflow as the ratio of the electromagnetic energy density 
to the baryon density is likely to be maximized above the event horizon. There 
is also probably a region between the inner disk and the event horizon in the equatorial 
plane where accreting gas falls 
supersonically onto the hole. This is likely to lie within the 
ergosphere where there exist
orbits of total negative energy.  In principle, magnetic field can 
drag matter onto these orbits and consequently extract spin energy from the hole 
without the magnetic field threading the 
event horizon although it is not clear if there is enough
time for this to happen in practice. (This is especially important if holes spin
at nearly their maximal rates as they will then diamagnetically exclude the flux.) 

Independent of the source of the power 
for ultrarelativistic jets, it is generally envisaged that jets are surrounded and 
collimated by a magnetic sheath derived from an orbiting accretion disk. Here 
the debate concerns whether the collimating magnetic field is basically large scale 
and organized, like that measured in the solar wind, or if it comprises 
disorganised loops that achieve the same effect. This hinges in turn on whether disks 
can maintain a large scale polarity in the open field lines that leave their surfaces,
as happens with the sun (and I believe to be the case), 
or whether there is no large scale poloidal field 
beyond that created stochastically by local, small scale instability. Theoretically,
I expect that numerical simulation will be of considerable interest. Observationally 
multiwavelength radio and mm monitoring is the key to understanding the
jet composition and drawing strong inferences concerning the jet origin.
\section{Telescopes and Space Missions}
I will conclude with a few remarks on the observational prospects.
\subsection{Small Colleges}
Many of the most variable AGN are quite bright and accessible to small telescopes
at good sites equipped with CCDs [Balonek, Nelson]. It is 
gratifying that this has enabled faculty at small colleges 
to involve enthusiastic undergraduates in invaluable monitoring efforts. I wonder
if there is an opportunity to widen this approach to include selected,
and strategically located international 
institutions. This could be an important educational outreach component of a space
facility proposal.
\subsection{XMM-Newton}
It was encouraging 
to hear that XMM-Newton is performing so well on orbit [O'Brien]. With its large
effective area, this should complement Chandra splendidly. It's optical monitor is 
ideal for AGN monitoring studies. Of course, its major contribution
will be to the unexpected, but we can anticipate even more accurate iron (and associated line)
profiles from nearby Seyferts and LINERs that will, I hope, assuage some of the doubts
that have been cast upon the ASCA results. I hope that it will also feasible to 
demonstrate the feasibility of X-ray reverberation studies. 
At the same time, it should produce
higher dispersion spectra, with superior signal to noise of the so-called warm absorption 
component which appears to be a large component of an AGN gas flow. In addition, it should 
contribute to X-ray blazar studies, not just by monitoring the continuum, but also by seeking 
broadened, high ionization state lines associated with outflowing gas, just like SS433.
\subsection{FUSE} 
The hopes for FUSE are no less [Kriss]. The major goal is surely to 
provide more accurate measurements
of the UV fluxes and resolve the arithmetical discrepancy 
in the ratio of the photoionizing input 
to the emission line output. Clearly absorption,
beaming and aspect are playing a role here, but it may 
be possible to remove these effects statistically by averaging over a large sample. 
I hope it will also contributed to the larger goals listed above by producing more 
accurate measurements of the quasar contribution to the metagalactic ionizing radiation
field. FUSE is also proving to be effective at measuring the high ionization
UV lines that originate from closest to the continuum source. In combination with X-ray
observations, it should help interpret the somewhat puzzling features of the
so-called warm absorber region.  Finally, it promises to address the ionization 
structure of the intergalactic medium, at least those phases with temperature of a few
times $10^5$~K, as discussed above.
\subsection{KRONOS}
A major reason for holding this meeting was to assess the scientific prospects and,
implicitly, the 
design criteria for KRONOS (Peterson, Horne, Collier). This is conceived as an explorer class 
space mission specialized to obtaining well sampled, multiwavelength continuum and line
emission from the near infrared to the soft X-ray bands from the brighter objects.
The sampling of the response functions that it promises will be far superior to anything
that currently exists and the mission seems ideally suited to making
``engineering drawings'' of AGN.       
\subsection{$\gamma$-rays}
The contribution of $\gamma$-ray observations over recent years to our understanding of AGN
has been much greater than most of us anticipated. With the unfortunate, though understandable,
demise of Compton Gamma Ray Observatory, it is important that AGILE (due for launch 
in 2003) and GLAST (2006) will continue the excellent discoveries of EGRET [Hartmann]
in the GeV 
energy range [Vercellone]. INTEGRAL due for launch in 2001, will tackle the 
instrumentally challenging MeV part of the spectrum (Paltarni). It should be able to measure 
the temperatures of Comptonizing coronae as well as search for annihilation lines.
Turning to even higher energies, VERITAS and its associated projects should be able
to monitor hundreds of nearby BL Lacs with their reported half hour variation
and probe the innermost regions of relativistic jets. They will also be able to perform an 
invaluable service for a larger community by measuring indirectly the intergalactic infrared
flux, a key tracer of star formation.
\subsection{Constellation-X and XEUS}
Perhaps the greatest long term hope for this field lies with the next generation of 
X-ray observatories [Mushotzky]. 
One of the major goals of Constellation-X is to carry out the 
reverberation mapping program of KRONOS at X-ray wavelengths. In this way it aspires to 
inspect the architecture of accretion disks as they approach the black hole ergosphere and,
most importantly, to produce 
indirect images that include the warped Kerr spacetime (Reynolds).

We are entering a new phase of our quest to make AGN as commonplace a part of 
astronomy as stars are today. What is clear is that, unlike stars, black holes 
need and use the total electromagnetic spectrum to get rid of the radiative 
by products of their nourishment. It is good that this spectral breadth is matched 
by the range of observational proposals that we have heard about here and, consequently,
we must work collectively to achieve our underlying goal of 
understanding how black holes accrete.
This was the rationale for and is the  principal message of this meeting.   
\section*{Acknowledgements}
I thank Brad Peterson for the invitation to attend this meeting,
his patience and for constructive critique
of the first draft of this summary.
Support under NSF contract AST 99-00866 is gratefully acknowledged.
\end{document}